\documentclass[twocolumn,showpacs,preprintnumbers,amsmath,amssymb]{revtex4}
\usepackage{graphicx}
\usepackage{dcolumn}
\usepackage{bm}
\begin{document}
\title{Generation of entangled channels for perfect teleportation channels using multi-electron quantum dots.}
\author{D. D. Bhaktavatsala Rao$^{1}$}
\author{Sayantan Ghosh$^{1}$}
\author{Prasanta K Panigrahi$^{1,2}$}
\affiliation{${}^1$Indian Institute of Science Education and Research Kolkata, INDIA.}
\affiliation{${}^2$Physical Research Laboratory, Navarangpura, Ahmedabad, INDIA.}

\date{\today}
\begin{abstract}
In this work we have proposed a scheme for generating $N$ qubit entangled states which can teleport an unknown state perfectly. By switching on the exchange interaction ($J$) between the qubits one can get the desired states periodically. A multi electron quantum dot can be a possible realization for generating such $N$ qubit states with high fidelity. In the limit of $N \rightarrow \infty$, there exists a unique time $t=\frac{1}{J}\cos^{-1}(-1/8)$ where the Hamiltonian dynamics gives the $N$ qubit state that can assist perfect teleportation. We have also discussed the effect of the nuclear spin environment on the fidelity of teleportation for a general $N$ qubit entangled channel.
\end{abstract}
\maketitle
\section{Introduction}
With the growing interest in teleporting unknown quantum states through entangled channels, many classes of $N$-qubit states have been studied. A few of these classes, for example the cluster states have been experimentally prepared \cite{natur}. Because of the large space in $N$-qubit Hilbert space one can always find quantum states which can be good for teleporting a specific kind of unknown states \cite{brown, illuminati}.The usefulness of these states are further characterized by implementing other protocols, for example the quantum state sharing (QSS)\cite{sreerampra}. Though many of the proposals have only a theoretical stand, very few received experimental importance. The robustness of various classes of $N$-qubit entangled states which can perform similar tasks can differ in the presence of noisy environments. In this work, we propose a scheme for generating $N$-qubit entangled channels in quantum dot systems where the coupling between electronic spins (qubits) is through the Heisenberg exchange interaction.

Spins of electrons in quantum dots are proposed to be good candidates for quantum computation \cite{divizenzo}. With the state-of-the-art technology in manufacturing and manipulating semiconductor nanostructures, quantum dots gives the promise for scalable quantum computing \cite{petta1,loss}. With a high level of control over the number of electrons\cite{ashoori}, and the applied external fields, desirable initial states can be achieved. The interaction between any two electronic spins can be controlled by the applied gate voltages \cite{petta1, petta2}.
The exchange interaction induced quantum gates, for example the ${SWAP}$ and $CNOT$ gates are implemented at picosecond time scales \cite{divizenzo}. The nuclear spins on the $2-D$ lattice give the dominant contribution to the decoherence of the electronic spin \cite{loss}. The time scales for this decay has been experimentally found to be of the order of a few nanoseconds \cite{fuji, petta2}. The effect of the spin environment really comes into the picture, either in repetitive usage of the dot, or, when there are large number of operations to be performed with these systems. The fidelity calculations show that one can perform $10^{2}$-$10^{3}$ gate operations before the real loss of spin polarization of the electrons.

In the step towards realizing scalable quantum computers, multi-electron quantum dots have been extensively studied both experimentally and theoretically in the past few years (see for example the review article\cite{hanson}). Three electron quantum dots arranged in a closed loop structure has been fabricated by Vidan \textit{et al.}\cite{vitan}. The experimental realization of multi-electron quantum dots and their usage for quantum computing was discussed thoroughly by Hanson \textit{et al.}\cite{hanson}.
In this paper we shall use Heisenberg exchange interaction between qubits to generate $N$-magnon entangled channels for teleporting an unknown quantum state perfectly. Multi-electron quantum dot systems can be a possible realization for the schemes that we propose in generating the desired entangled states.

\section{Teleportation of one qubit states}
Teleportation of an unknown state through an entangled two-qubit channel proposed by Bennett \textit{et al.}\cite{bennett} has been demonstrated using entangled photons \cite{phototele}, atomic states \cite{atomtele}, spin states in Nuclear Magnetic Resonance \cite{nmrtele} and other solid state systems \cite{qdottele}. The usage of multi-particle entangled states for implementing a similar protocol was studied initially using three and four qubit $GHZ$ states \cite{karl}. In addition to the $N$-qubit $GHZ$ states, $W$ states form another class of entangled states, and are well studied for their high symmetry. $W$ states can be generated from any $N$-qubit interactions which conserve the $\hat{z}$ component of the total spin of qubits.

The symmetric three qubit $W$ state,$|\psi \rangle =\frac{1}{\sqrt{3}}\Big[|100\rangle+|010\rangle+|001\rangle \Big]$
fails to teleport the unknown state perfectly to Bob. In this connection, Agarwal and Pati \textit{et al.} \cite{pati} have proposed a modification of the $W$ state which can teleport the unknown state perfectly. Though these states can teleport single qubits perfectly, they cannot be used for quantum state sharing. Later it was shown that the modified $W$ states can be connected to $GHZ$ states by performing entangling operations on any two qubits \cite{liu}. Since Alice is in possession of two qubits, depending on the task given to her she can either use her two qubits to teleport one qubit perfectly, or can perform appropriate unitary transformations on her qubits (converting it to $GHZ$ state) and use it for quantum state sharing by giving one of her qubits to Charlie \cite{qss}. Thus the modified $W$ state can be a key source for quantum protocols as different tasks can be performed in accordance with the requirement. This state has been subject to extensive study in the past few years. Proposals using cavity QED experiments for generating these states have been given \cite{chinese}. Considering the importance of these states, we propose an experimental way of generating these states using exchange interaction between the qubits in quantum dot systems, and further generalize to $N$ qubits.

Let us consider a $3$-qubit state which is an eigen state of the total $\hat{S^z}= \sum_{i=1}^3S^z_i$ operator given by,
\begin{equation}
\label{eq:3qu}
|W_{3}\rangle =\alpha_{1}|100\rangle +\alpha_{2}|010\rangle+\alpha_{3}|001\rangle,
\end{equation}
where, $|\alpha_1|^2+|\alpha_2|^2+|\alpha_3|^2=1$. If one of the qubits, say the last one is given to Bob, and the other two are in Alice's possession, then, Alice can perfectly teleport a one qubit state $|\psi\rangle =a|0\rangle+b|1\rangle$ to Bob when the coefficients satisfy the following relation given by,
\begin{equation}
\label{eq:cond_3_1}
|\alpha_3|^2=|\alpha_1|^2+|\alpha_2|^2.
\end{equation}
A simple state which satisfies the above condition is,
\begin{equation}
\label{eq:w3}
|W_3\rangle=\frac{1}{2}\Big[|100\rangle+|010\rangle+\sqrt{2}|001\rangle\Big].
\end{equation}
With a more general parametrization, one can show that, $\alpha_1=\frac{1}{\sqrt{2}}$, $\alpha_2=\frac{1}{\sqrt{2}}\sin\phi\rm{e}^{\iota \chi_1}$ and $\alpha_3=\frac{1}{\sqrt{2}}\cos\phi\rm{e}^{\iota \chi_1}$, where, $0 \le \phi \le 2\pi$ and similarly, $0\le \chi_1\rm{ , }\chi_2\le  2\pi$. On the contrary, if Bob is given the first qubit in the state $|W_3\rangle$, then for perfect teleportation,
\begin{equation}
\label{eq:cond_3_2}
|\alpha_1|^2=|\alpha_2|^2+|\alpha_3|^2.
\end{equation}
Conditions given in Eq. \eqref{eq:cond_3_1} and Eq. \eqref{eq:cond_3_2} lead to completely different states. One can see that the imbalance in the weights of the basis states $|1\rangle$ and $|0\rangle$ for each qubit is responsible for such conditions. To overcome these imbalances, one can construct a $W$-like state,
\begin{equation}
\label{eq:w3_like}
|\tilde{W}_3\rangle=\frac{1}{2}\Big[|100\rangle+|010\rangle+|001\rangle+|111\rangle\Big],
\end{equation}
where the basis states of all qubits are on equal footing. The problem with the above state is that it does not conserve the total $\hat{S_z}$, and hence becomes difficult to generate using simple interactions.
\begin{figure}[htb]
\begin{center}
   \includegraphics[width=10.0cm]{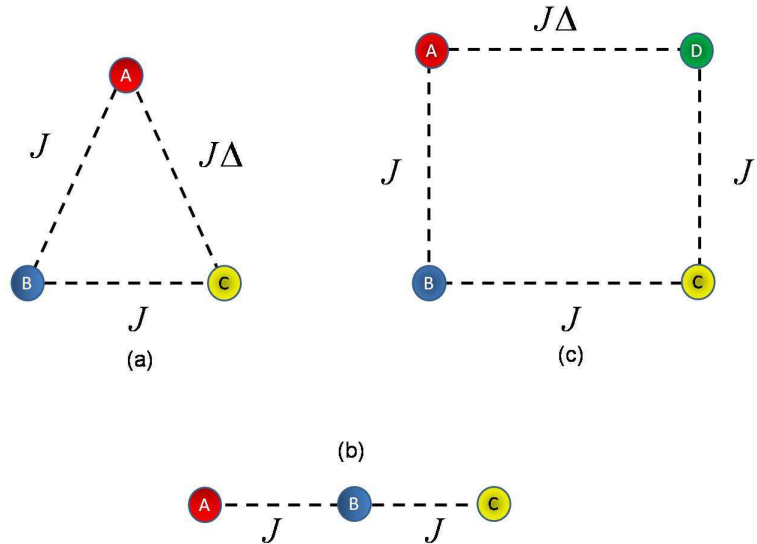}
\end{center}
 \caption{Representation of closed and open chains for $3$ and $4$ qubits interacting through Heisenberg exchange interaction. For $\Delta=1$ the chain is perfectly closed and $\Delta=0$ the chain is open.}
 \end{figure}
\subsection{Generation of the $|W_3\rangle$ state}
The three qubit exchange interaction can be described by the following Hamiltonian,
\begin{equation}
\label{eq:ham}
\mathcal{H}=J\vec{S}_A.\vec{S}_B+J\vec{S}_B.\vec{S}_C+J\Delta \vec{S}_A.\vec{S}_C,
\end{equation}
where, $J$ is the strength of the interaction. The exchange interaction is typically of the order of $0.01\rm{eV}$ in strongly interacting systems. The parameter $\Delta$ determines the closeness of the $3$-qubit chain. If $\Delta=1$, this is a perfectly closed chain and for $\Delta=0$, it is an open chain. We shall show how different kinds of $|W_3\rangle$ states can be generated by varying this parameter $\Delta$.

The above Hamiltonian can be straight forwardly diagonalized, and the time evolution of various $3$-qubit states can be found. Since we are interested only in one magnon states, we shall start with an initial $3$-qubit state, $|\psi(0)\rangle =|100\rangle$. Since the Hamiltonian given in 
Eq. \eqref{eq:ham} conserves the $\hat{z}$ component of the total spin, i.e., $[\mathcal{H},\sum_{k}S_{k}^{z}]=0$ where, $k=A,B,C$, the state at a later time will be
\begin{equation}
|\psi(t)\rangle=\alpha_1(t)|100\rangle+\alpha_2(t)|010\rangle+\alpha_3(t)|001\rangle,
\end{equation}
where the time dependent coefficients are given by

\begin{eqnarray}
\alpha_1(t) &=& \frac{1}{6}\left[2{\rm e}^{-itE_1} + 3{\rm e}^{-itE_2} + {\rm e}^{-itE_3}\right], \nonumber \\
\alpha_2(t) &=& \frac{1}{3}\left[{\rm e}^{-itE_1} - {\rm e}^{-itE_3}\right], \nonumber \\
\alpha_3(t) &=& \frac{1}{6}\left[2{\rm e}^{-itE_1} -{\rm e}^{-itE_2} + {\rm e}^{-itE_3}\right]. \nonumber \\
\end{eqnarray}
The eigenvalues $E_1 = J(2+\Delta)/4, E_2 = _3J/4, E_3 = (\Delta - 4)/4$.
The condition for perfect teleportation when the first qubit is given to Bob, Eq. \eqref{eq:cond_3_2} for the above state gives,
\begin{equation}
\label{eq:nl}
3\cos(Jt\frac{1+2\Delta}{2})+\cos\frac{3Jt}{2}+\frac{3}{2}\cos((1-\Delta)Jt)-1 = 0.
\end{equation}
\begin{figure}[htb]
\begin{center}
   \includegraphics[width=8.0cm]{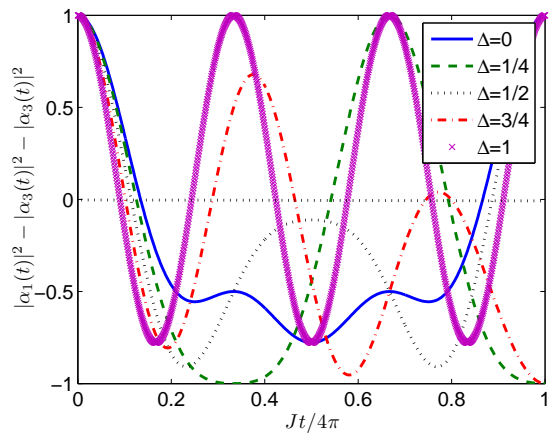}
\end{center}
 \caption{(Color online)We have plotted Eq. (\ref{eq:nl}) as a function of time. The times at which the curves intersect with the horizontal line are the solutions to the non-linear equation given in Eq. (\ref{eq:nl}). The asymmetry parameter $\Delta$ is varied from $0$ (open) to $1$ (perfectly closed). The times at which one can find solutions to Eq. (\ref{eq:nl}) increases with decreasing $\Delta$. }
 \end{figure}

One can now extract the times for which the above equation is satisfied. If the interaction between the qubits is switched off at those times, the three qubit state obtained will be the one which can be used as a perfect teleportation channel. The question of how exactly can one switch off the interaction depends on further details of the experiment which we do not discuss here. Finding the roots for the non-linear equation given in Eq. \eqref{eq:nl} is in general non-trivial for arbitrary values of $\Delta$. For the case of open and closed chains, one can find the roots to the above equation.

For perfectly closed chains corresponding to $\Delta=1$, Eq. \eqref{eq:nl} yields the solution which  has a periodicity of $4\pi/3$, given by
\begin{equation}
\label{eq:sol_nl_1}
Jt=\frac{2}{3}\cos^{-1}\Big(-\frac{1}{8}\Big).
\end{equation}
In addition to this one also finds that $|\alpha_1(t)^2|=\frac{1}{2}$ and $|\alpha_2(t)|^2=|\alpha_3(t)|^2=\frac{1}{4}$, at the above given time. The three qubit state at the above given time is simply given by,
\begin{equation}
|\psi(t=\tau)\rangle=\frac{1}{2}\Big[\sqrt{2}\rm{e}^{\iota\phi_1}|100\rangle+\rm{e}^{\iota\phi_2}|010\rangle+\rm{e}^{\iota\phi_2}|001\rangle\Big],
\end{equation}
where $\tau=\frac{2}{3}[2n\pi+\cos^{-1}(-\frac{1}{8})]$, and $\phi_1=\rm{tan}^{-1}(-\sqrt{2})$ and $\phi_2=\rm{tan}^{-1}(\frac{\sqrt{2}}{3})$. Other than the phases $\phi_1$ and $\phi_2$, the above state is given by Pati \textit{et al.}\cite{pati}. Since we are interested to give the first qubit to Bob, we have started with the initial state $|100\rangle$. Instead, if the last qubit have to be given to Bob, we would have stated with the initial state $|001\rangle$.
The solutions to Eq. \eqref{eq:nl} for other values $\Delta$ can be seen from Fig.2. As $\Delta$ decreases from one to zero, the time scale shifts towards  higher values. Although the times for which the solution exists changes with $\Delta$, the probability of finding the state in $|100\rangle$ is still one half i.e., $|\alpha_1|^2 = 1$.
Hence for any value of $\Delta$ one can still write the final three qubit state when the condition given in Eq. $\eqref{eq:nl}$ is satisfied, as
\begin{equation}
|\psi(t=\tau_\Delta)\rangle=\frac{1}{2}\Big[\sqrt{2}\rm{e}^{\iota\phi_1(\Delta)}|100\rangle+\rm{e}^{\iota\phi_2(\Delta)}|010\rangle+\rm{e}^{\iota\phi_3(\Delta)}|001\rangle\Big].
\end{equation}

\subsection{A N-qubit teleporting channel with one magnon}
In this section we shall consider the case of a $N$ qubit entangled channel having one magnon excitation. In teleporting a single qubit, states with at least one magnon excitation are required. To construct a $n$ magnon state, first consider the ground state of the $N$ qubit system to be $|0\rangle = |00\cdots 0\rangle$ and now flip $n$ spins at various sites. This can be in done in ${}^N C_n$ ways. Now the linear superposition of all such states represent a $n$ magnon state given by
\begin{eqnarray}
|N;n\rangle = \sum_{i_1i_2.......i_n = 1}^{N}C_{i_1i_2\cdots i_n}|i_1\cdots i_7\cdots i_n\rangle,
\end{eqnarray}
where $\sum_{i_1i_2.......i_n = 1}^{N}|C_{i_1i_2\cdots i_n}|^2 = 1$. In the above $ |i_1\cdots i_7\cdots i_n\rangle$ represents the $N$ qubit state with spins flipped at positions $i_n$. We shall consider all the complex coefficients to be some phase factors only. For $N=2$ the above state becomes $|2;1\rangle = C_1|01\rangle + C_2|10\rangle$ and for $N=3$, $|3;1\rangle = C_1|100\rangle + C_2|010\rangle + C_3|001\rangle$. Note that $|3;2\rangle$ state is just the mirror reflection of the state $|3;1\rangle$. For two magnon excitations the state must at least consist of $N=4$ qubits.

Since we are interested to teleport the information of a single qubit we shall give one qubit to Bob and keep the remaining $N-1$ qubits with Alice. The initial state of the total system is given by
\begin{eqnarray}
|\psi\rangle_i = (\alpha|0\rangle + \beta|1\rangle)\otimes\sum_{i=1}^N C_i |i\rangle,
\end{eqnarray}
We shall break up the channel into Alice and Bob qubits separately as follows $\sum_{i=1}^N C_i |i\rangle = \sum_{i=1}^{N-1} C_i |i\rangle|0\rangle +C_N|00\cdots 0\rangle|1\rangle$.
We shall rewrite the initial state such that Alice's qubits and Bob qubit gets factorized into four different products, given by
\begin{eqnarray}
|\psi\rangle_i &=& \frac{1}{2}\bigg[(|\xi_1\rangle + |\xi_4\rangle)(\alpha|0\rangle + \beta|1\rangle) +(|\xi_1\rangle - |\xi_4\rangle)(\alpha|0\rangle - \beta|1\rangle) \nonumber \\
&+& (|\xi_2\rangle + |\xi_3\rangle)(\beta|0\rangle + \alpha|1\rangle)+(|\xi_2\rangle - |\xi_4\rangle)(-\beta|0\rangle + \alpha|1\rangle)\bigg] \nonumber
\end{eqnarray}

where $|\xi_1\rangle = |0\rangle\sum_{i=1}^{N-1} C_i |i\rangle$, $|\xi_4\rangle = |1\rangle|00\cdots 0\rangle$, $|\xi_2\rangle = |0\rangle|00\cdots 0\rangle$ and $|\xi_3\rangle = |1\rangle\sum_{i=1}^{N-1} C_i |i\rangle$.
For perfect teleportation Alice should perform a measurement on a orthonormal basis which would imply that the states
$|\xi_1\rangle \pm |\xi_4\rangle$ and $|\xi_1\rangle \pm |\xi_4\rangle$ should be mutually orthogonal. This immediately imposes the condition that 
\begin{eqnarray}
\sum_{i=1}^{N-1} |C_i|^2 = |C_N|^2.
\end{eqnarray}
 Considering the complex coefficients $C_i$ to be only phase factors we immediately see that the $N$ qubit channel which can teleport a single qubit state is given by
\begin{eqnarray}
\label{onemst}
\hspace{-5mm}
|\psi\rangle_C = \frac{1}{\sqrt{2(N-1)}}\sum_{j=1}^{N-1}{\rm e}^{i2\pi j/N}|j\rangle|0\rangle \pm \frac{1}{\sqrt{2}}|00\cdots 0\rangle|1\rangle.
\end{eqnarray}
For $N=2$ this is the usual Bell state and for $N=3$ this is the modified $W$ state discussed in the earlier section. Note that deterministic teleportation using the above class of states can be possible only for $N=2$, as Alice has the complete orthonormal basis for measurement. Similar states were reported earlier in \cite{liu}.

The above analysis can be generalized to find out the $n$ magnon states which can be used for perfectly teleporting one qubit state, given by
\begin{widetext}
\begin{eqnarray}
|\psi\rangle_C = \frac{1}{\sqrt{2\left[{}^{N-1}C_n\right]}}\left[\sum_{j_1j_2..j_n=1}^{N-1}|j_1j_2\cdots j_n\rangle|0\rangle \pm \sqrt{\frac{N-n}{n}}\sum_{j_1j_2..j_{n-1}=1}^{N-1}|j_1j_2\cdots j_{n-1}\rangle|1\rangle\right],
\end{eqnarray}
\end{widetext}
where ${}^{N-1}C_n = (N-1)!/n!(N-n-1)!$. By adding appropriate phase factors the above state can be much more generalized. As one can see from the above that only when $N=2$, Alice can perform a complete orthonormal basis measurement.

\subsection{Generation from Hamiltonian dynamics}
In this section we shall construct these $N$ qubit teleporting channels from switching on the exchange interaction between the qubits. Here, we shall consider only closed chains. The interaction Hamiltonian is simply given by
\begin{eqnarray}
\mathcal{H} = J\sum_i \vec{S}_i \cdot \vec{S}_{i+1},
\end{eqnarray}
with the cyclic periodic condition $N+1 = 1$. The one magnon states which are eigenstates of the above Hamiltonian are $|k\rangle = \sum_n {\rm e}^{ikn}|n\rangle$, where $k=2\pi\lambda/N,~\lambda = 0,1,2\cdots N-1$ and $|n\rangle$ represents the site number at which the spin is flipped. The eigenvalues corresponding to these $k$ states are given by $E_k = J(1-\cos k)$, where the ground state energy is set to zero.

Starting with the initial state $|\psi(0)\rangle = |100\cdots 0\rangle$ (first qubit goes to Bob), the state at any later time can be straightforwardly found, which is given by
\begin{eqnarray}
\label{cond2}
|\psi(t)\rangle = \sum_k \sum_n {\rm e}^{iE_kt}e^{ik(n-1)}|n\rangle.
\end{eqnarray}
Now imposing the condition for perfect teleportation one finds
\begin{eqnarray}
\label{nmcond}
\left|\sum_k {\rm e}^{iE_kt}\right|^2 - \sum_{n=2}^{N} \left |\sum_k {\rm e}^{iE_kt} e^{ik(n-1)} \right|^2 = 0.
\end{eqnarray}
Instead of solving the above equation, we can just solve for the times when the left hand side of the above equation becomes equal to half, i.e., $\left|\sum_k {\rm e}^{iE_k(t=\tau)}\right|^2 = 1/2$.
For finite $N$ we can still find that $\tau = \frac{2}{3J}\cos^{-1}\Big(-\frac{1}{8}\Big)$. For large $N$ we can replace the summation in the above equation by an appropriate integral given by
\begin{eqnarray}
\sum_{p=0}^{N-1}{\rm e}^{iJt\cos(2\pi p/N)}\approx \frac{N}{2\pi}\int_0^{2\pi} dx {\rm e}^{iJt\cos x} = N\mathcal{J}_0(Jt),
\end{eqnarray}
where $\mathcal{J}_0(Jt)$ is the zeroth order Bessel function. Using this, the condition for the existence of solution to Eq.\ref{cond2} is then simply given by $\mathcal{J}_0(Jt) = 1/\sqrt{2}$. In Fig.2 we have plotted Eq.\ref{cond2} for various values of $N$. From the nature of the Bessel function one can see that there will no revival of $|\alpha_1(t)|^2$ close to $1/2$ as $N$ becomes large. Hence there can exist only one solution in the large $N$ and that time is precisely given by $\tau$ defined in Eq.(\ref{{eq:sol_nl_1}}) and the state at this time will be similar to the one given in Eq.(\ref{onemst}). Thus we have found a more general condition of generating the $N$ qubit entangled channels to teleport one qubit perfectly across a chain of non interacting quantum dots.
\begin{figure}[htb]
\begin{center}
   \includegraphics[width=8.0cm]{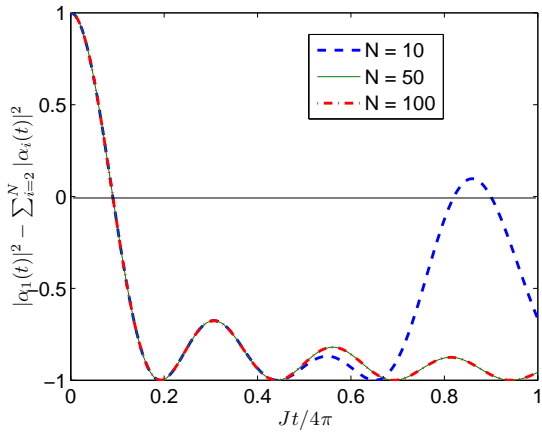}
\end{center}
 \caption{(Color online) We have plotted Eq. \ref{nmcond} as a function of time for different values of $N$. The times at which the curves intersect with the horizontal line are the solutions to the non-linear equation given in the above equation. In large $N$ limit, there are no strong revivals indicating that there can exist only one solution corresponding to the first intersection with the horizontal line. Though for finite $N$ there will always be revivals, the revival time increase drastically with $N$.}
 \end{figure}
\section{Decoherence from nuclear spin environments}
The performance of every quantum protocol is limited by the dissipating effects of environments. In quantum dot systems, interaction of the electron spin with nuclear spin bath gives the dominant contribution to its decoherence. Since the $N$ qubit entangled states described above are generated at the order of few pico seconds and then separated, only the effects of individual environments for each spin need to be considered. Hence the system bath interaction is simply given by
\begin{eqnarray}
\mathcal{H}_{SE} = \sum_{ij}K_{i} \vec{S}_i \cdot \vec{I}_i.
\end{eqnarray}
In the above equation $\vec{I}_i = \sum_k \vec{I}_{i,k}$ represents the total spin of the nuclear spin bath for the $i^{th}$ quantum dot. The dynamics for the above Hamiltonian can be solved exactly \cite{durga}. Considering the initial state of the bath to be completely unpolarized, the time evolution of each of the $N$ qubit teleporting channels can be found from
\begin{eqnarray}
\rho_N(t) &=& {\rm Tr}_{I_1, I_2 \cdots I_N}\sum_{i,j}\bigg\lbrace\left(a_i(t) + b_i(t)\vec{S}_i\cdot\vec{I}_i\right)\nonumber \\&&~~~~~~~~~~~|\psi_N\rangle\langle\psi_N|\otimes\rho_B(0)\nonumber
~\left(a^*_j(t) + b^*_j(t)\vec{S}_j\cdot\vec{I}_j\right)\bigg\rbrace,
\end{eqnarray}
where $a_i = \cos \Lambda_{i}t +iK_{i}\sin \Lambda_{i}t /2\Lambda_{i}$ and $b_i =2iK_{i}\sin \Lambda_{i}t/\Lambda_{i}$,  where $2\Lambda_{i} = K_{i}(I_{i}+1/2)$.

In studying decoherence, density matrix representation of the multi-qubit state can be quite convenient. For example 
the three qubit teleporting channel derived earlier in Sec-IIA,
$|\psi\rangle=\frac{1}{2}\Big[\sqrt{2}|100\rangle+\rm{e}^{\iota\phi}|010\rangle+\rm{e}^{\iota\phi}|001\rangle\Big]$can be rewritten in terms of the spin operators corresponding to each qubit as
\begin{eqnarray}
\rho &=& \frac{1}{8}\bigg[\mathcal{I} - (S^z_B + S^z_C)(\mathcal{I}+2S^z_A) - \sqrt{2}{\rm e}^{i\phi}(S^+_AS^-_B +S^+_AS^-_C)\nonumber \\ && + S^+_BS^-_C 
-2\sqrt{2}{\rm e}^{i\phi}(S^+_A+S^-_BS^z_C + S^+_AS^z_BS^-_C) \nonumber \\ &&-2S^z_AS^+_BS^-_C + 8S^z_AS^z_BS^z_C + h.c\bigg],
\end{eqnarray}
where $S^{\pm} = S^x\pm i S^y$.
In time the spin operators of each qubit become time-dependent and the state at any later time obtained after tracing out the bath degrees of freedom will have the same form given above, with only the spin operators replaced with their time-dependnet forms, given by
\begin{eqnarray}
\vec{S}_i(t) = \vec{S}_i(0)\left\lbrace\frac{1}{3}+\frac{2}{3}\left(1-\frac{t^2}{\tau_i^2}\right){\rm e}^{-t^2/2\tau_i^2}\right\rbrace,
\end{eqnarray}
where $\tau_i = \frac{{2}}{K_i\sqrt{N_i}}$, where $N_i$ are the number of nuclear spins in the $i^{th}$ quantum dot.

The third particle correlations for example, $S^z_AS^x_bS^y_C$ decay much faster (with a decay rate $\tau_1 + \tau_2 + \tau_3$), in comparison to the two particle correlations, $S^x_AS^y_B$ which decay at the rate of $\tau_1 + \tau_2$. 
Though the state has lost all its three particle correlations it can be still entangled because of the two particle correlations which decay comparatively slower. In contrast, the $GHZ$ state which has only three particle correlations,  becomes completely mixed quickly indicating that the state is not good even for probabilistic teleportation. For the above described environmental interactions the modified $W$ states can be quite useful for teleporting unknown states with better fidelity than $GHZ$ states.
The time evolution for a general $N$ qubit state can be written following the above analysis. The higher order correlations decay faster, and the multi-qubit states become mixed faster with increasing $N$, thereby decreasing the fidelity of teleportation.  To overcome this situation one needs to develop control schemes where the entanglement of multi-qubit states can be preserved for longer times.

In conclusion we have shown how exchange interaction among spin-$1/2$ particles can be used to generate a class of multi qubit entangled channels that can be used for teleportation. For times $t=\frac{1}{J}\cos^{-1}(-1/8)$ we obtain the $N$ qubit entangled state which can be used for teleportation and further for QSS under certain conditions. For a typical quantum dot system this time scale can be of the order of picoseconds.
For smaller number of qubits these states can be generated periodically, whereas for large $N$ there exists a unique time where one can obtain such entangled states. In the presence of local decoherence for each qubit higher order correlations decay faster due to which the modified states show promise for better fidelity than $GHZ$ states which have higher order polarizations only. 
\begin{acknowledgments}
One of the authors Rao, would like to acknowledge the financial support provided by the J. C. Bose fellowship.
\end{acknowledgments}

\end{document}